\documentclass[12pt]{iopart}

\usepackage{iopams} 
\usepackage{bm,graphics,graphicx,epsfig,color,hyperref}

\newcommand{\ud}{\mathrm{d}}
\newcommand{\ui}{\mathrm{i}}

\begin{document}

\title[Gravitational-wave recoil]{Gravitational-Wave Recoil from the Ringdown Phase of Coalescing Black Hole Binaries}

\author{Alexandre Le Tiec$^{1}$, Luc Blanchet$^1$ and Clifford M. Will$^{2,1}$}

\ead{letiec@iap.fr, blanchet@iap.fr, cmw@wuphys.wustl.edu}
\address{$^1$ $\mathcal{G}\mathbb{R}\varepsilon{\mathbb{C}}\mathcal{O}$, Institut d'Astrophysique de Paris --- UMR 7095 du CNRS, Universit\'e Pierre et Marie Curie, 98$^{\rm bis}$ boulevard Arago, 75014 Paris, France \\ $^2$ McDonnell Center for the Space Sciences, Department of Physics, Washington University, St. Louis MO 63130 USA}

\begin{abstract}
The gravitational recoil or ``kick'' of a black hole formed from the merger of two orbiting black holes, and caused by the anisotropic emission of gravitational radiation, is an astrophysically important phenomenon. We combine (i) an earlier calculation, using post-Newtonian theory, of the kick velocity accumulated up to the merger of two non-spinning black holes, (ii) a ``close-limit approximation'' calculation of the radiation emitted during the ringdown phase, and based on a solution of the Regge-Wheeler and Zerilli equations using initial data accurate to second post-Newtonian order. We prove that ringdown radiation produces a significant ``anti-kick''. Adding the contributions due to inspiral, merger and ringdown phases, our results for the net kick velocity agree with those from numerical relativity to 10--15 percent over a wide range of mass ratios, with a maximum velocity of $180$ km/s at a mass ratio of $0.38$.
\end{abstract}


\maketitle

\section{Introduction and summary}

The gravitational recoil of an isolated system in response to the anisotropic emission of gravitational radiation (sometimes also called the ``kick'') is a phenomenon with potentially important astrophysical consequences \cite{Me.al.04}. One of the most intriguing is the possibility that a massive black hole formed from the inspiral and merger of two progenitor black holes could receive enough of a kick to displace it from the center of the galaxy where the merger occurred, or to eject it entirely from the galaxy. This could affect the growth history of massive black holes \cite{Sc.07}. Observational evidence for such a kicked black hole has even been reported \cite{Ko.al.08}.

The calculation of such kicks within general relativity has been carried out in a variety of ways. Earlier analytic or semi-analytic estimates of the gravitational recoil include a perturbation calculation (valid for small mass ratios) during the final plunge \cite{Fa.al.04}, a post-Newtonian calculation valid during the inspiraling phase together with a treatment of the plunge phase \cite{Bl.al.05}, an application of the effective-one-body formalism \cite{DaGo.06}, and a close-limit calculation with Bowen-York type initial conditions \cite{So.al.06}. 
 
Following recent advances in numerical calculations of binary black holes \cite{Pr.05,Ca.al.06,Ba.al.06}, the problem of gravitational recoil received considerable attention from the numerical relativity community. These computations led to increasingly accurate estimates of the kick velocity from the merger along quasicircular orbits of black holes without spin \cite{Ca.05,Ba.al3.06,He.al.07,Go.al.07,Go.al.09}, and with spin \cite{He.al2.07,Ko.al.07,Ca.al.07}; from head-on collisons \cite{Ch.al.07}; and from hyperbolic orbits \cite{He.al.09}. In particular these numerical simulations showed that very large kick velocities can be obtained in the case of spinning black holes for particular spin configurations. Nevertheless, as the very detailed multipolar analysis of the binary black hole recoil by Schnittman {\em et~al.} \cite{Sc.al.08} illustrates, analytic and/or semi-analytic methods are still very useful for gaining more physical understanding of the relaxation of binary black holes to their final equilibrium state.

In the simplest case of unequal mass, non-spinning black hole binaries on quasicircular orbits, the kick velocity as a function of time shows a very distinctive pattern \cite{Ba.al3.06,Go.al.07}: the recoil increases monotonically during the inspiral and plunge phases up to a maximum around the onset of merger, and then decreases quickly to a final asymptotic value, as much as 30 percent smaller than the maximum. This braking occurs during the phase where the newly formed black hole emits gravitational radiation in a superposition of quasinormal ``ringdown'' modes, and is known as the ``anti-kick''. For a reduced mass parameter value $\eta \equiv m_1 m_2/(m_1 +m_2)^2 \simeq 0.19$, the value for which the kick is a maximum, the peak value is around $250$ km/s while the final kick is around $175$ km/s.  

Building on previous work based on a multipolar post-Minkowskian formalism \cite{BlDa.86,Bl.95,Bl.98}, Blanchet, Qusailah and Will \cite{Bl.al.05} (hereafter BQW) derived the linear momentum flux from compact binaries at second post-Newtonian (2PN) order beyond the leading effect. BQW augmented their 2PN estimate of the recoil up to the innermost circular orbit (ICO) by integrating the resulting 2PN-accurate flux from the ICO down to the horizon on a plunge geodesic of the Schwarzschild geometry. They found that the recoil monotonically increases as the plunge progresses. Within their error bars, the resulting recoil was found to agree well with the maximum value of the kick velocity as calculated by numerical relativity up to the onset of the anti-kick. And indeed the maximal kick velocity in numerical computations occurs more or less at a separation of roughly $2M$, where $M=m_1+m_2$ is the total mass, as inferred from the times at which the maximum kicks were found to occur in \cite{Ba.al3.06,Go.al.07} (such an inference cannot be made rigorously, of course, but does coincide roughly with where other numerical diagnostics indicated the onset of the merger). At this point, the BQW computation ended for lack of a method to evaluate the contribution from the subsequent ringdown phase. This paper reports the results of incorporating such a method.

\begin{figure}[t]
	\begin{center}
		\begin{tabular}{c}
		\hspace{-0.5cm}\includegraphics[width=6.5cm,angle=-90]{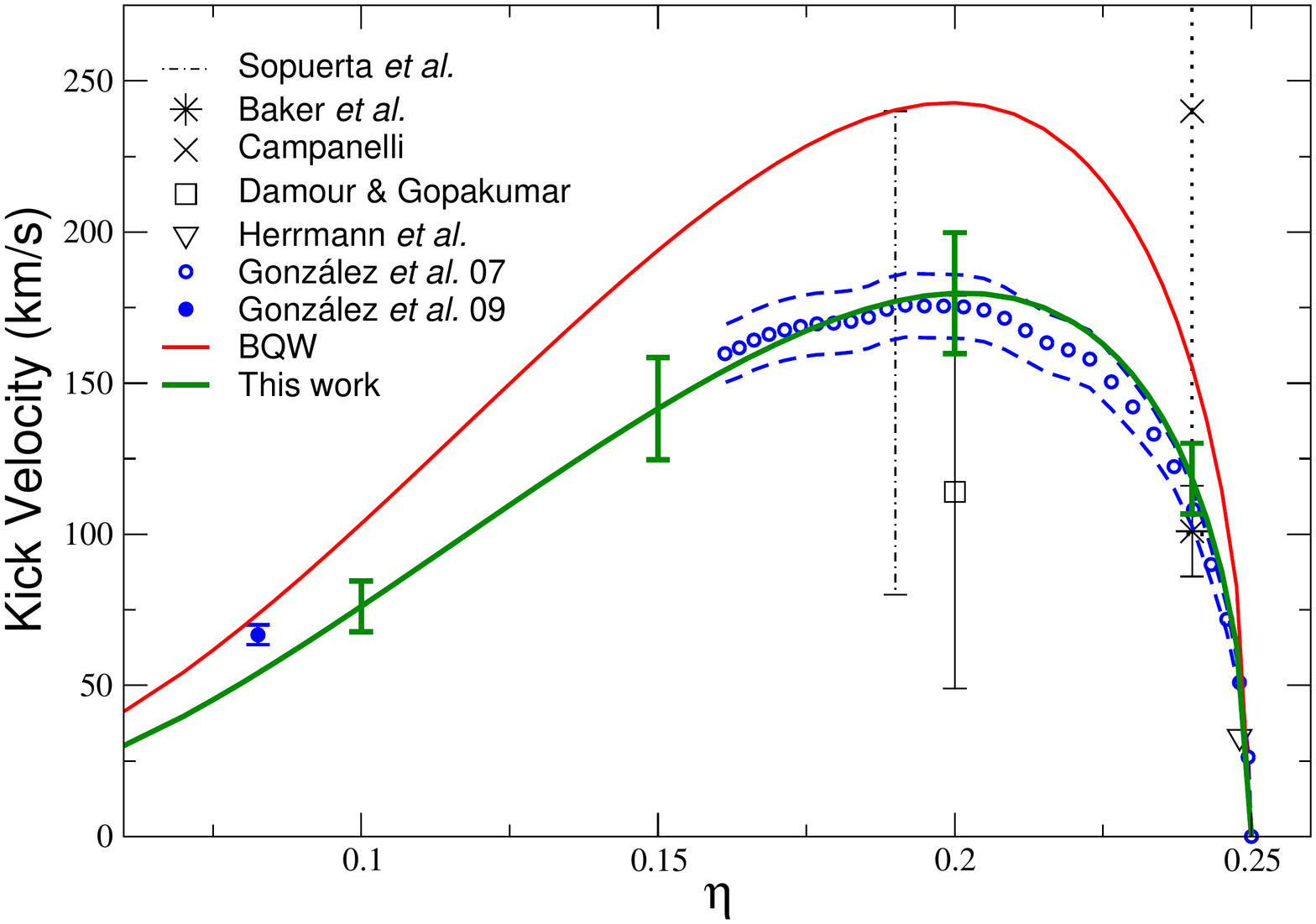}
		\hspace{-0.9cm}\includegraphics[width=6.5cm,angle=-90]{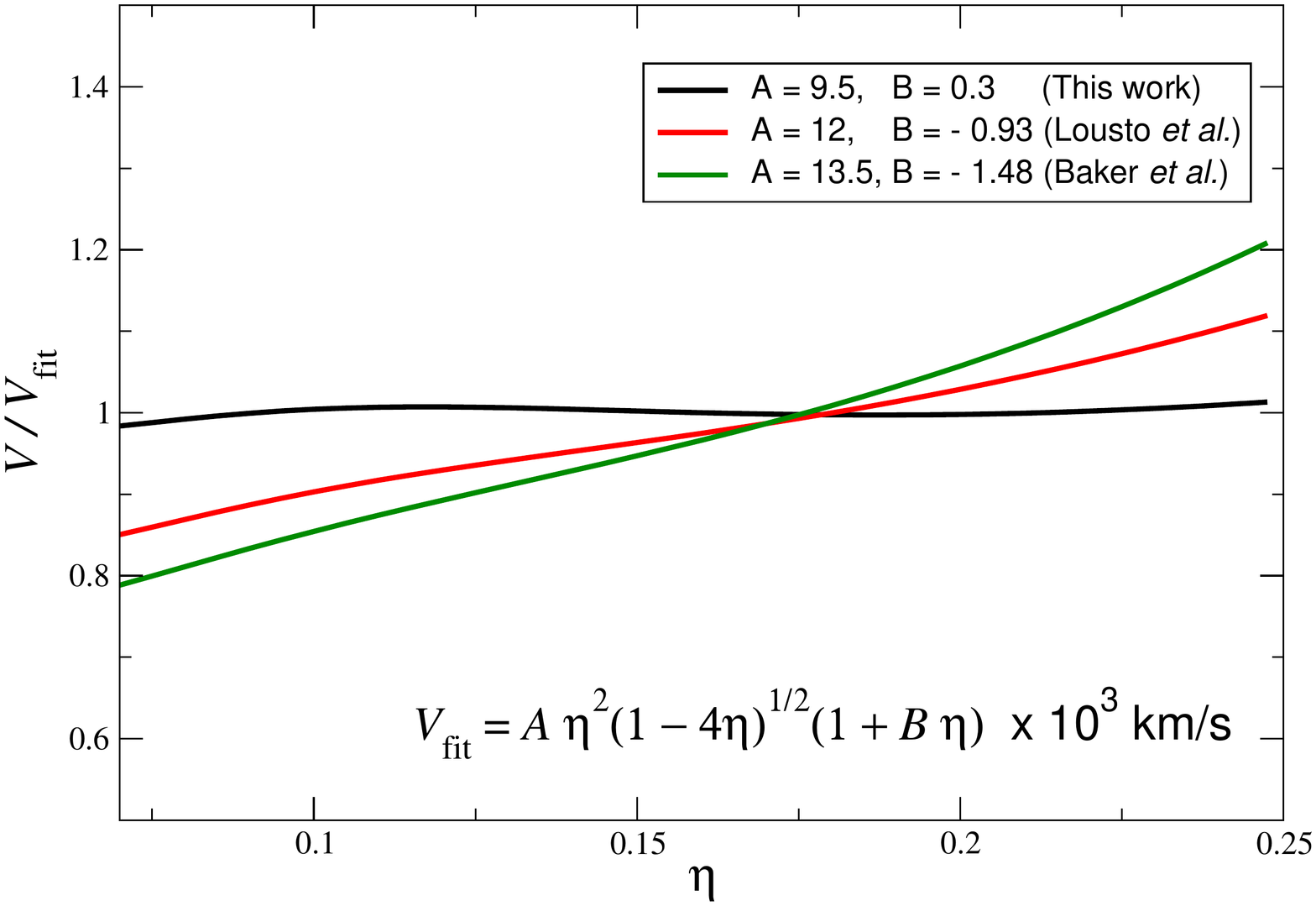}
		\end{tabular}
		\caption{Left panel: Comparison with numerical calculations \cite{Ca.05,Ba.al3.06,He.al.07,Go.al.07,Go.al.09} and other analytic or semi-analytic methods \cite{Bl.al.05,DaGo.06,So.al.06}. Right panel: Comparison with fitting formulas derived from this paper (black) and from numerical relativity results \cite{Ba.al.08,Lo.al.09}.}
		\label{fig:sperhake}
	\end{center}
\end{figure}
Le Tiec and Blanchet \cite{LB09} have developed a ``close-limit approximation'' (CLA) for black hole binaries that uses 2PN-accurate initial conditions. In this 2PN-CLA framework, the 2PN metric for two bodies in close proximity is recast as a perturbation of a Schwarzschild black hole. The resulting perturbation is then used as initial data to evolve numerically the Regge-Wheeler and Zerilli equations in order to calculate the gravitational radiation emitted subsequently. The purpose of the present paper is to use the resulting waveforms to compute the recoil generated during the ringdown phase. By adding vectorially the 2PN-CLA results to the 2PN results of BQW for the inspiral plus merger, we prove that the effect of the ringdown on the recoil is indeed to produce an anti-kick, and we find that the total kick generated by the inspiral plus merger plus ringdown phases is in good agreement with numerical computations for non-spinning black hole binaries.
 
Our central results are shown in Fig.~\ref{fig:sperhake}. In the left panel, the curve with error bars (green in the color version) shows our combined kick velocities, with the error bars estimated by varying the radius at which the 2PN and CLA methods are matched. The top curve (red) is the kick from the BQW pure 2PN calculation up to the merger. The sequence of dots and accompanying dashed lines (blue) are from an exhaustive series of numerical simulations by Gonz\'alez {\em et al.} \cite{Go.al.07,Go.al.09}. In this figure are also shown individual points and error estimates from some earlier analytic or semi-analytic estimates such as \cite{DaGo.06,So.al.06}, as well as other numerical computations.

A number of authors have fit the kick velocity to the empirical formula \cite{Bl.al.05,Go.al.07}
\begin{equation}
V_{\rm fit} = A \,\eta^2 (1-4\eta)^{1/2}\,(1 + B\,\eta) \times 10^3 \, {\rm km/s} \,.
\end{equation}
The leading $\eta^2 (1-4\eta)^{1/2}$ dependence derives from the lowest-order, or ``Newtonian'' calculation \cite{Fi.83,FiDe.84}. A fit to our 2PN-CLA results yields $A= 9.5$ and $B = 0.3$. The right panel of Fig.~\ref{fig:sperhake} shows the ratio of our kick velocities to this fitting formula (black), together with similar ratios to fitting formulas derived from numerical relativity \cite{Ba.al.08,Lo.al.09}. It can be seen that our kick velocities are systematically higher than those from numerical relativity in the equal-mass limit, and systematically lower in the small mass-ratio limit. However, we find it striking that in the regime where the kick velocity is substantial, say between $\eta = 0.08$ and $\eta = 0.24$, our 2PN-CLA calculation agrees with numerical relativity to 10--15 percent, and confirms the intuition that ringdown radiation generates an anti-kick that partially offsets the kick accumulated during the plunge.

The remainder of this paper is organized as follows: In Sec.~\ref{secII} we describe briefly the 2PN and CLA methods for calculating the gravitational recoil from the inspiral, plunge and ringdown phases (full details will be presented elsewhere \cite{LBW09}), and we describe the numerical implementation of the method and some of the checks and diagnostics performed. Concluding remarks are made in Sec.~\ref{secIII}. We use geometrical units $G=c=1$.

\section{Gravitational recoil from the inspiral, plunge and ringdown phases}
\label{secII}

The flux of linear momentum carried away by gravitational waves from a general isolated system can be written in terms of the gravitational-wave polarization states $h_+$ and $h_\times$ as \cite{MTW}
\begin{equation}\label{Pi}
	\frac{ \ud P^i}{\ud t} = \lim_{r \to +\infty} \left\{ \frac{r^2}{16 \pi} \oint n^i \, \vert \dot{h}_+ - \ui \, \dot{h}_\times \vert^2 \, \ud \Omega \right\} ,
\end{equation}
where the dot stands for a partial time derivative, and $\ud \Omega$ is the solid angle associated with the direction of propagation $n^i$. 

BQW \cite{Bl.al.05} expanded the waveforms in terms of radiative multipole moments using the post-Minkowski framework, and expressed them in terms of source multipole moments of mass-type and current-type, valid to 2PN order, including contributions of ``tails''. Restricting to binary systems on quasicircular inspiral orbits, they obtained $\ud P^i/\ud t$, and after integrating with respect to time, dividing by the total mass $M$ and changing sign, obtained the recoil velocity 
\begin{eqnarray}
\label{recoilx}
V^i &=& \frac{464}{105}\,\eta^2\, (1-4\eta)^{1/2} x^4\left[1+\left(-\frac{452}{87} -\frac{1139}{522}\eta\right)\,x
  +\frac{309}{58}\,\pi\,x^{3/2}\right.\nonumber\\ &&\qquad\left.
  +\left(-\frac{71345}{22968}+\frac{36761}{2088}\eta
  +\frac{147101}{68904}\eta^2\right)\,x^2\right]\, u^i \, ,
\end{eqnarray}
where $x=(M\omega)^{2/3}$, $\omega$ being the orbital angular frequency, and $u^i$ is the unit vector directed from the less massive toward the more massive body. (See \cite{Ra.al.09} for a generalization including spin effects.) This formula gives the kick velocity $V^i_{\rm inspiral}$ accumulated during inspiral up to the ICO (defined by $x_{\rm ICO} = \frac{1}{6}$).

Starting with $V^i_{\rm inspiral}$, which is always a small contribution to the total kick, BQW then integrated the 2PN expression for $\ud P^i/\ud t$ along a plunge orbit of a Schwarzschild black hole of mass $M$ from the ICO down to a radius of order $2M$. A key to that step was to change integration variable from coordinate time $t$, which is singular on the event horizon, to a ``proper angular frequency'' variable $\bar{\omega}=\ud\psi/\ud\tau$ of the plunge orbit, which is regular on the horizon. The resulting net kick was the data for the curve labeled BQW (red) plotted in Fig.~\ref{fig:sperhake}. Here we repeat this calculation, except that we terminate the plunge integration at a Schwarzschild coordinate radius $r_{\rm match}$, whose value is chosen to lie between $2M$ (the minimum allowed by the method) and $2.5M$. At this radius we match the 2PN kick (namely $V^i_{\rm inspiral}+V^i_{\rm plunge}$), to the result of our 2PN-CLA calculation, to which we now turn. Later we will test the sensitivity of the final result to the value of the matching radius $r_{\rm match}$.

The idea of the 2PN-CLA method \cite{LB09} is to take the spacetime metric for a binary system accurate to 2PN order, where the two bodies are on a quasicircular orbit of initial separation $r_{12}$ (in Schwarzschild-like coordinates), which is assumed to be of order $2M$. The metric is then re-expanded in powers of $r_{12}$, resulting in a Schwarzschild metric of a black hole of mass $M$ plus correction terms that vanish in the limit $\eta \to 0$. Carrying out a multipolar expansion, one can identify the $(\ell, m)$ components of the Regge-Wheeler and Zerilli functions used in black-hole perturbation theory.

It is possible to express the linear momentum flux (\ref{Pi}) in terms of the Regge-Wheeler and Zerilli functions $\Psi_{\ell,m}^{({\rm e,o})}$ as defined in Eqs.~(5.1) of \cite{LB09}. The waveform takes the form
\begin{equation}\label{h_psi}
	h_+ - \ui \, h_\times = \frac{1}{r} \sum_{\ell,m} \sqrt{\frac{(\ell+2)!}{(\ell-2)!}} \left( \Psi_{\ell,m}^{({\rm e})} + \ui \, \Psi_{\ell,m}^{({\rm o})} \right) {}_{-2}Y_{\ell,m} + \mathcal{O}(r^{-2}) \, ,
\end{equation}
where the superscripts $({\rm e})$ and $({\rm o})$ denote even and odd-parity respectively, the summations on the integers $\ell, m$ range from 2 to infinity for $\ell$, and from $-\ell$ to $\ell$ for $m$, and where ${}_{-2}Y_{\ell,m} (\theta,\varphi)$ are the spin-weighted spherical harmonics of spin $-2$ \cite{NePe.66,Go.al.67}. More details on this standard result are given in the Appendix of \cite{LB09}, including its less well-known generalization at any order in $r^{-1}$. We insert Eq.~(\ref{h_psi}) into Eq.~(\ref{Pi}) and find (see \cite{Ru.al.08} for details)
\begin{equation}\label{P+X}
		\fl \frac{\ud P_x}{\ud t} + \ui \, \frac{\ud P_y}{\ud t} = - \frac{1}{8 \pi} \sum_{\ell,m} \left[ \ui \, a_{\ell,m} \, \dot{\Psi}_{\ell,m}^{({\rm e})} \dot{\bar{\Psi}}_{\ell,m+1}^{({\rm o})} + b_{\ell,m} \biggl( \dot{\Psi}_{\ell,m}^{({\rm e})} \dot{\bar{\Psi}}_{\ell+1,m+1}^{({\rm e})} + \dot{\Psi}_{\ell,m}^{({\rm o})} \dot{\bar{\Psi}}_{\ell+1,m+1}^{({\rm o})} \biggr) \right] ,
\end{equation}
where $a_{\ell,m} = 2 (\ell-1)(\ell+2) \sqrt{(\ell-m)(\ell+m+1)}$ and $b_{\ell,m} = \frac{(\ell+3)!}{(\ell+1)(\ell-2)!} \sqrt{\frac{(\ell+m+1)(\ell+m+2)}{(2\ell+1)(2\ell+3)}}$, and the overbar denotes complex conjugation. Because of the symmetry with respect to the orbital plane, we naturally find $\ud P_z / \ud t = 0$. This would no longer remain true if we were to include spin-orbit coupling terms for spinning black holes in the initial PN metric. The master functions $\Psi_{\ell,m}^{({\rm e,o})}$ obey the wave equations
\begin{equation}\label{ZRW}
	\left( \partial^2_t - \partial^2_{r_*} + \mathcal{V}_\ell^{({\rm e,o})} \right) \Psi_{\ell,m}^{({\rm e,o})} = 0 \, ,
\end{equation}
where the tortoise coordinate $r_*$ is related to the Schwarzschild radial coordinate $r$ by
$r_* = r + 2M \, {\rm ln}( {r}/{2M} - 1 )$, and where the potentials $\mathcal{V}_\ell^{({\rm e,o})}$ are given by
\begin{equation}
	\mathcal{V}_\ell^{({\rm e,o})} = \left( 1 - \frac{2M}{r} \right) \! \left( \frac{\ell (\ell +1)}{r^2} - \frac{6M}{r^3} \, \mathcal{U}_\ell^{({\rm e,o})} \right) ,
\end{equation}
with $\mathcal{U}_\ell^{({\rm e})} = \frac{\lambda_\ell (\lambda_\ell+2) r^2 + 3M (r-M)}{(\lambda_\ell \, r + 3M)^2}$ and $
\mathcal{U}_\ell^{({\rm o})} = 1$, where $\lambda_\ell = \frac{1}{2}(\ell-1)(\ell+2)$.

\begin{figure}
	\begin{center}
		\includegraphics[width=7cm,angle=-90]{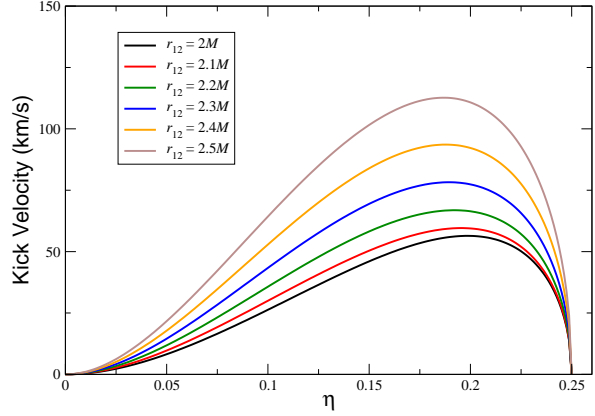}
		\caption{Magnitude of the recoil velocity $V_{\rm ringdown}$ generated during the ringdown phase as a function of the symmetric mass ratio $\eta$ for different initial separations $r_{12}$.}
		\label{fig:recoil}
	\end{center}
\end{figure}
We evolve the equations (\ref{ZRW}) with 2PN-accurate initial conditions computed with an initial separation $r_{12}$, and developed in the CLA as described above; detailed inputs are Eqs. (3.6)--(3.7) and (4.4)--(4.5) of Ref.~\cite{LB09}
(see also \cite{So.al.06} for an alternative CLA calculation using
different initial conditions). Then, inserting the numerically generated master functions $\Psi_{\ell,m}^{({\rm e,o})}$ into Eq. (\ref{P+X}), we calculate the flux of linear momentum up to octupolar order during the ringdown phase, and, integrating that with respect to time, dividing by $M$ and changing the sign, we obtain the ringdown contribution $V^i_{\rm ringdown}$ to the total kick. Fig.~\ref{fig:recoil} shows the magnitude $V_{\rm ringdown} = |V^i_{\rm ringdown}|$  as a function of $\eta$ for various values of $r_{12}$.

\begin{figure}
	\begin{center}
		\begin{tabular}{c}
		\hspace{-0.5cm}\includegraphics[width=6.5cm,angle=-90]{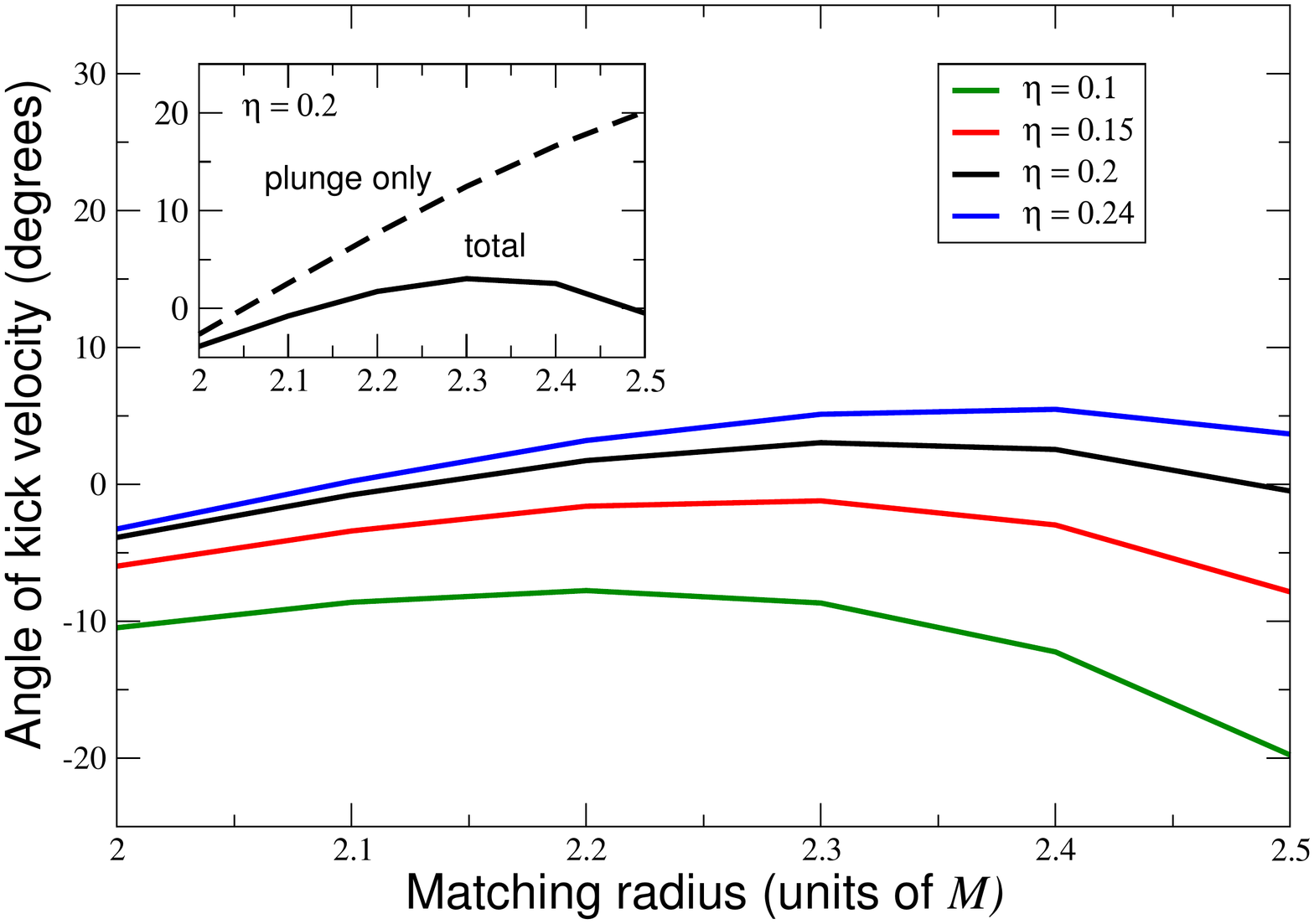}
		\hspace{-0.9cm}\includegraphics[width=6.5cm,angle=-90]{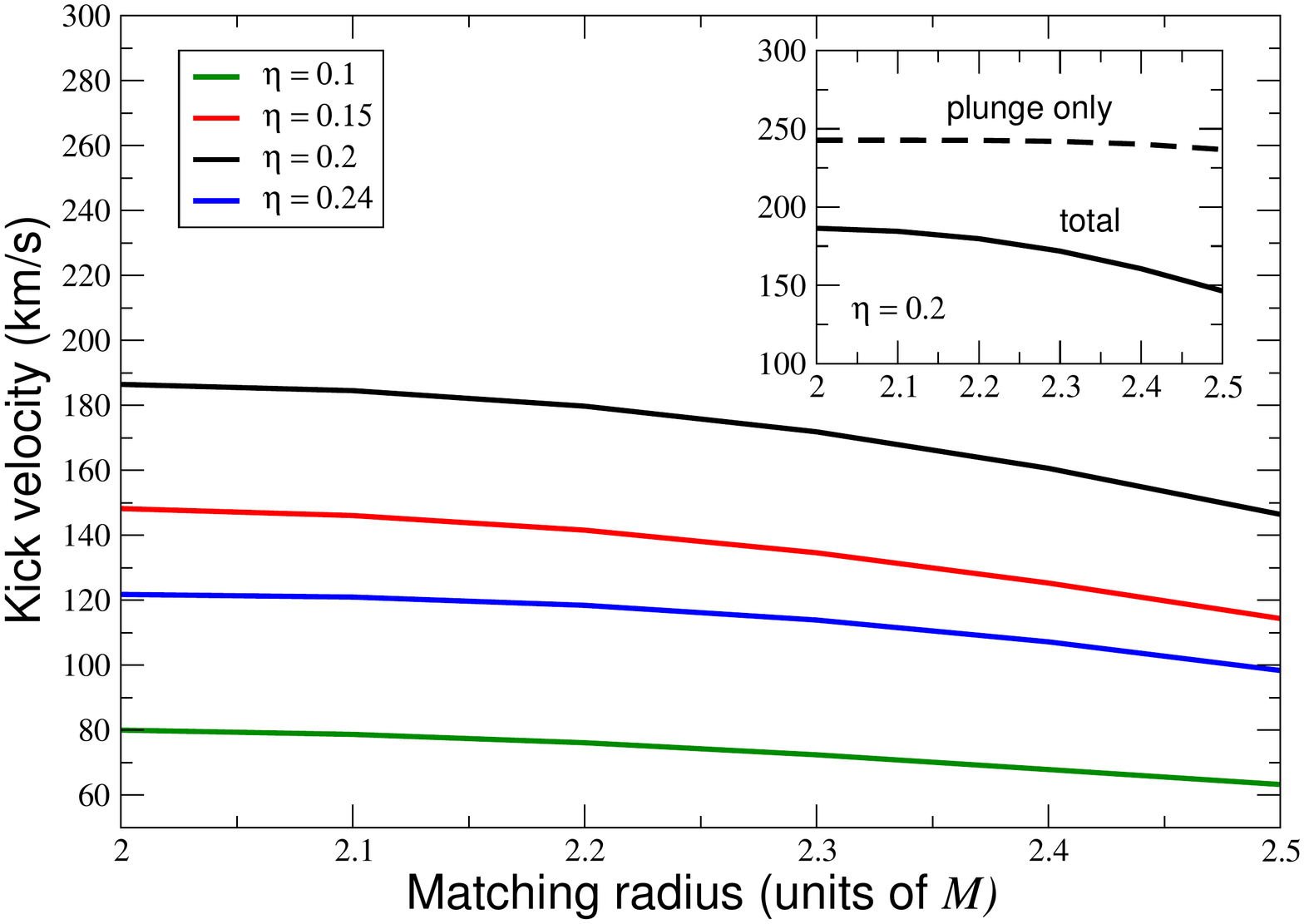}
		\end{tabular}
		\caption{Dependence of the result for the direction (left panel) and magnitude (right panel) of the recoil velocity upon the matching radius $r_{\rm match}$.}
		\label{fig:match}
	\end{center}
\end{figure}
Choosing $r_{12} = r_{\rm match}$, we then add up vectorially the results for the inspiral, plunge and ringdown phases to obtain $V^i = V^i_{\rm inspiral} + V^i_{\rm plunge} + V^i_{\rm ringdown}$. In all cases we find that the direction of the ringdown kick is approximately opposite to that of the accumulated inspiral plus plunge kick. Not surprisingly, the direction or phase of the inspiral plus plunge kick is sensitive to the radius $r_{\rm match}$ at which the 2PN calculation terminates. But, most satisfactorily, when we add the ringdown kick, the final direction is relatively insensitive to the value of $r_{\rm match}$, as shown in the left panel of Fig.~\ref{fig:match} (see especially the inset panel). Similarly, the right panel shows that the magnitude of the total kick velocity is also only weakly sensitive to $r_{\rm match}$. In Fig.~\ref{fig:sperhake}, we have chosen $r_{\rm match} = 2.2M$ as being a value where the phase and magnitude of the kick do not vary too much, and estimated error bars by varying $r_{\rm match}$ between $2M$ and $2.5M$. Unlike BQW, we have not attempted to estimate errors caused by the neglect of higher PN corrections in the CLA method.

\section{Conclusions}\label{secIII}

We have found that the recoil velocity of coalescing, non-spinning black holes can be calculated using a combination of post-Newtonian theory for the inspiral and plunge, and a close-limit approximation for the ringdown, with results that agree closely with those from full-scale numerical relativity. We have also used this method to determine the total energy and angular momentum radiated during inspiral, plunge and ringdown; details will be published elsewhere \cite{LBW09}. An obvious, though non-trivial next step would be to incorporate the effects of spins in this approach.

\ack
We wish to thank Ulrich Sperhake for providing the data necessary to produce the left panel of Fig.~\ref{fig:sperhake}, on which we have superimposed our results. This work was supported in part by the National Science Foundation, Grant No.\ PHY 06--52448, the National Aeronautics and Space Administration, Grant No.\ NNG-06GI60G, the Programme International de Coop\'eration Scientifique of the Centre National de la Recherche Scientifique (CNRS--PICS), Grant No. 4396, and the McDonnell Center for the Space Sciences. CMW acknowledges the hospitality of the Institut d'Astrophysique de Paris, where this work was completed. 

\section*{References}

\bibliographystyle{iopart-num}
\bibliography{/tmp_mnt/netpapeur/users_home4/letiec/Publications/ListeRef}

\end{document}